# Electrooptic response of chiral nematic with oblique helicoidal director

Jie Xiang[1], Sergij V. Shiyanovskii[1], Corrie Imrie[2], Oleg D. Lavrentovich[1,*]

1 Liquid Crystal Institute and Chemical Physics Interdisciplinary Program,
Kent State University, Kent, USA
2 Department of Chemistry, University of Aberdeen, AB24 3UE Scotland, UK

**Abstract.** Electrically induced reorientation of liquid crystals (LCs) is a fundamental phenomenon widely used in modern technologies. We observe experimentally an electro-optic effect in a cholesteric LC with a distinct oblique-helicoidal director deformation. The oblique helicoid, predicted in late 1960-ies, is made possible by recently developed dimer materials with an anomalously small bend elastic constant. Theoretical, numerical, and experimental analysis establishes that both the pitch and the cone angle of the oblique helicoid increase as the electric field decreases. At low fields, the oblique helicoid with the axis parallel to the field transforms into a right-angle helicoid (the ground state of field-free cholesteric) with the axis perpendicular to the field.

Nematic liquid crystals (LCs) are orientationally ordered fluids with average orientation of molecules described by the so-called director $\hat{\mathbf{n}}$. In the simplest case of the uniaxial nematic (N), the ground state is $\hat{\mathbf{n}}$=const. When some or all of the nematic molecules are chiral, the director twists in space, following a right-angle helicoid; the structure is called a cholesteric (or chiral nematic) N*. The electrooptical effects of practical significance are based on the electrically-induced reorientation of $\hat{\mathbf{n}}$ caused by the dielectric anisotropy. The field-induced modification of N* helicoid is typically of the two types: changing the pitch of the right angle-helicoid [1] (for example, in diffractive elements [2]) or realigning the helicoid axis as the whole (which is used in bistable displays [3]). In these effects, the fundamental character of the right-angle helicoidal twist remains intact. Almost 50 years ago, R.B. Meyer [4] and P. G. de Gennes [5] predicted that there should exist a very distinct mode of electrically induced deformation, with the director forming an oblique helicoid. This "heliconical" state was never proven to exist experimentally, mainly because its existence requires the bend elastic constant $K_3$ to be much smaller than the twist constant $K_2$, a condition that is not satisfied in typical nematics formed by rod-like molecules. In this work, we demonstrate the existence of the oblique helicoidal state in a cholesteric LC formed by recently developed molecular dimers, in which the flexible aliphatic chain that links two rigid rod-like arms makes the ratio $\kappa = K_3/K_2$ anomalously small [6-8]. We expand the theory by establishing (a) the field dependence of the cone angle and (b) the scenario of the oblique-to-right-angle helicoid transformation. We present experimental and numerical data on the field dependencies of pitch and cone angle and establish the transition scenarios between the uniaxial nematic state at high fields, oblique helicoid at intermediate fields, and N* right-angle helicoid at low fields.

The field-induced director deformations are described within the framework of the Frank-Oseen free energy functional. Neglecting the effects of electric field non-locality, the energy density for a left-handed N* writes

$$f = \frac{1}{2}K_1(\nabla \cdot \hat{\mathbf{n}})^2 + \frac{1}{2}K_2(\hat{\mathbf{n}} \cdot \nabla \times \hat{\mathbf{n}} - q_0)^2 + \frac{1}{2}K_3(\hat{\mathbf{n}} \times \nabla \times \hat{\mathbf{n}})^2 - \frac{1}{2}\Delta\varepsilon\varepsilon_0(\hat{\mathbf{n}} \cdot \mathbf{E})^2, \quad (1)$$

where $K_1$ is the splay elastic constant, $q_0 = 2\pi/P_0$, $P_0$ is the pitch of N*, $\varepsilon_a = \varepsilon_\parallel - \varepsilon_\perp > 0$ is the local dielectric anisotropy, representing the difference between the permittivity parallel and perpendicular to $\hat{\mathbf{n}}$; $\mathbf{E}$ is the applied electric field. In absence of the field, the ground state is a right-angle left-handed helicoid $\hat{\mathbf{n}} = (\cos q_0 y, 0, \sin q_0 y)$. Suppose that the field is applied along the $x$-axis, $\mathbf{E} = (E, 0, 0)$. When the field is very high, the director is parallel to it, $\hat{\mathbf{n}} = (1, 0, 0)$, because $\varepsilon_a > 0$. Suppose now that the field is reduced, so that the tendency to twist caused by chiral nature of molecules, can compete with the dielectric torque. Below some threshold field [4] $E_{NC} = \frac{2\pi}{P_0} \frac{K_2}{\sqrt{\varepsilon_0 \varepsilon_a K_3}}$, the unwound nematic transforms into a heliconical state with the director that follows an oblique left-handed helicoid, $\hat{\mathbf{n}} = (\cos\theta, \sin\theta \sin\varphi, \sin\theta \cos\varphi)$ with the cone angle $\theta > 0$ and the angle of homogeneous azimuthal rotation $\varphi(x) = 2\pi x / P$. The heliconical pitch $P$ is inversely proportional to the field [4]:

$$P = \frac{2\pi}{E} \sqrt{\frac{K_3}{\varepsilon_0 \varepsilon_a}} = \frac{\kappa E_{NC} P_0}{E}. \tag{2}$$

Minimization of the free energy functional based on the density (1) also relates the cone angle $\theta$ to the strength of the electric field:

$$\sin^2 \theta = \frac{\kappa}{1-\kappa}\left(\frac{E_{NC}}{E} - 1\right). \tag{3}$$

Since $\varepsilon_a > 0$, it is clear, however, that the cone angle would not increase continuously to its limiting value $\theta = \pi/2$ as that would mean $\hat{\mathbf{n}}$ being perpendicular to $\mathbf{E}$ everywhere. One should thus expect a complete reorganization of the oblique helicoid with an axis along $\hat{\mathbf{x}} \| \mathbf{E}$ into a right-angle helicoid with the axis *perpendicular* to $\mathbf{E}$, at fields lower than:

$$E_{N*C} \approx E_{NC} \frac{\kappa\left(2 + \sqrt{2(1-\kappa)}\right)}{1+\kappa}. \tag{4}$$

The last expression is derived by balancing the energies of the right-angle and oblique helicoidal states with a small $\kappa$ in the external field. With this theoretical background, we now consider the experimental situation.

**Materials and techniques/cells.** We used a LC dimer material 1",7"-bis(4-cyanobiphenyl-4'-yl)heptane ($NC(C_6H_4)_2(CH_2)_7(C_6H_4)_2CN$, CB7CB) which shows a uniaxial N phase between 116°C and 103°C with a positive dielectric anisotropy [9], sandwitched between the isotropic and the twist-bend nematic phase $N_{tb}$ [7,16]. We measured the dielectric permittivities parallel and perpendicular to the director as $\varepsilon_\parallel = 7.3$ and $\varepsilon_\perp = 5.9$, respectively; the elastic constants were determined by the Frederiks transition technique [1] to be $K_1 = 5.7$ pN and $K_2 = 2.6$ pN. All data correspond to 106°C. To prepare the N* phase, we doped CB7CB with a small amount (1wt%) of chiral (left-handed) dopant S811. The phase diagram is different from the case of an un-doped CB7CB: N* melts into an isotropic fluid at $T_{N*I} = 112°C$ and transforms into a homochiral version of $N_{tb}$ at $T^* = 99°C$. The pitch $P_0$ of the N* phase, measured in the Grandjean-Cano wedge [1], decreases from 8.8 μm at $T^*+1°C$ to 6.2 μm at $T_{N*I} -1°C$, see Supplement. The electro-optic experiments were performed at the temperature $T^*+3°C$, at which $P_0 = (7.5 \pm 0.5)$ μm.

We used flat glass cells of thickness $d =$(11-16) μm. The glass substrates were coated with polyimide PI2555 that sets a homeotropic (perpendicular) orientation of the molecules. When the cell is filled with N*, it shows a fingerprint texture with the helicoid axis in the plane $(x,y)$ of the cell. This geometry allows one to clearly visualize the periodic structure of both the heliconical and cholesteric structures, as the wave-vector of director modulations in both cases is confined to the plane $(x,y)$. To assure a uniform alignment of the helicoid, the polyimide coatings were rubbed unidirectionally along the axis $x$. For the polarizing optical microscopy (POM) study, two aluminum foil electrodes were placed between the glass plates to apply the electric field parallel to the rubbing direction $x$. The distance between the electrodes was 140 μm. For optical diffraction and optical retardance mapping by PolScope [10], the cells with patterned indium tin oxide (ITO) electrodes on one of the substrates were used, and the distance between the electrodes was $L = 100$ μm. The AC field of frequency 3 kHz was used to explore the scenarios of structural transformations of the N* cells. Because of the cell geometry and in-plane arrangement of the electrodes, the electric field is inhomogeneous, being somewhat larger near the electrodes. To establish the spatial pattern of the electric field, we use COMSOL Multiphysics finite-element based solver, see Supplement. The simulations show that in the central part of the cell, the field is uniform and horizontal in the middle of the cell. For example, for the ITO case, for the applied voltage $U = 100$ V, the field is 0.7 V/μm with a 5% accuracy in the range $-20\,\mu m \leq x \leq 20\,\mu m$ and across the entire extension of the LC slab along the $z$-direction. The field acting in the center of the cell can be calculated as $E = \beta U / L$, where $\beta$ is the correction coefficient, determined by numerical simulations to be 0.67 for ITO and 0.75 for aluminum electrodes.

**Experimental results.** We start the experiment at a high field, 4 V/μm, at which the helical structure of N* is completely unwound with $\hat{\mathbf{n}} \parallel \mathbf{E}$ (except possibly in narrow regions near the glass plates because of the homeotropic anchoring), showing no periodic modulations, Fig.1(a). The texture is dark when viewed between two crossed polarizers, one of which is parallel to $\mathbf{E}$. When the field is slowly decreased, the texture starts to brighten at $E_{NC} = (1.1 \pm 0.07)$ V/μm, showing a periodic modulation along the $x$-axis, Fig.1(b). The period increases with the decrease of the electric field, Fig.2(a). The effect is not transient, as for a fixed field, the structure relaxes to feature a well-defined period. Adjustment of the period to the varied electric field is achieved by nucleation and propagation of edge dislocations of Burgers vector equal to the single period of the structure. As the field is decreased further, at some other threshold $E_{N*C} = (0.35 \pm 0.07)$ V/µm, the structure changes completely, by nucleating regions with the wave-vector of periodic modulation that is perpendicular to $\mathbf{E}$, Fig.1(c). The new structure has a much larger period of about 9 µm, which decreases as the electric field becomes smaller.

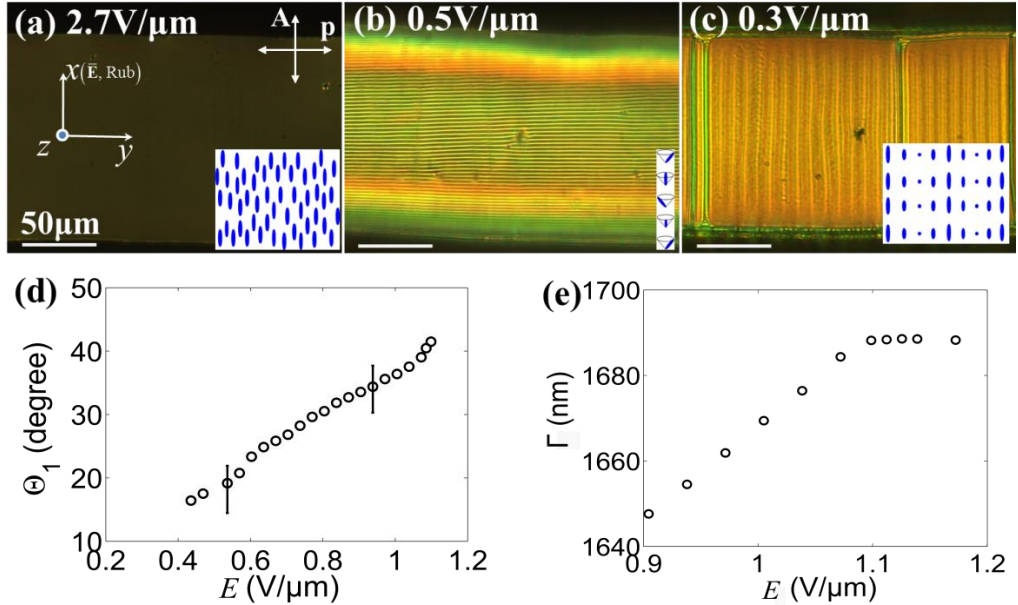

FIG.1. Electric-field induced (a) unwound nematic with the director parallel to the electric field; (b) heliconical state with the director following an oblique helicoid with the axis along the electric field, and (c) right-angle helicoid state of the cholesteric, as seen under the polarizing optical microscope. All scale bars are 50µm. (d) First order diffraction angle of the heliconical state as a function of the applied electric field. (e) Optical phase retardance as a function of the applied electric field in the vicinity of nematic-to-heliconical transition.

The described scenario corresponds to the transition from the nematic to oblique helicoid structure at $E_{NC}$, with a subsequent oblique-to-right angle helicoid first-order transformation with axis reorientation at $E_{N*C} < E_{NC}$. To demonstrate the oblique helicoidal state in the range $E_{N*C} < E < E_{NC}$, we use optical diffraction and PolScope. Optical diffraction experiment is performed with a He-Ne laser beam (λ=633nm) director normally to the cell. Polarization of incident light is varied by a rotating polarizer. The diffraction pattern is projected onto a screen 9.5 cm away from the sample. The heliconical state is a polarization-sensitive phase diffraction grating. For normal incidence, the diffraction condition is $m\lambda = P\sin\Theta_m$, where $m$ is the diffraction order, $\Theta_m$ is the corresponding diffraction angle. For small cone angles $\theta$, the first-order diffraction intensity $\propto \sin^2 2\theta$ is expected to be higher than the second-order diffraction intensity $\propto \sin^4 \theta$. This is indeed the case, as the values of $P$ calculated from the field dependence of $\Theta_1$, Fig.1(d), match the POM data very well, Fig.2(a).

The field dependence $P(E)$ follows closely the theoretically expected behavior [4] $P \propto 1/E$, Fig.2(a), which allows one to extract an important information on the elastic constants of N*. According to Eq.(2), $\kappa = EP/E_{NC}P_0$, which yields $\kappa = K_3/K_2 \approx 0.12$ with the experimental data on $P(E)$, $E_{NC}$ and $P_0 = 7.5\,\mu m$. The smallness of $\kappa$ satisfies the restrictions imposed by Meyer-de Gennes theory [1,4]. Moreover, the experimental $E_{N*C} = (0.35 \pm 0.07)$ V/$\mu$m agrees with the value $E_{N*C} = 0.39$ V/$\mu$m obtained from Eq.(4) when $\kappa = 0.12$. The twist modulus is independently calculated from the definition of $E_{NC}$ as

$$K_2 = \varepsilon_0 \varepsilon_a \kappa \left(\frac{P_0 E_{NC}}{2\pi}\right)^2 = 2.6\,\text{pN},$$

the same as measured in the N phase 106°C. With the above data, one deduces a rather small value of the bend elastic constant in N*, $K_3 = 0.3\,\text{pN}$.

PolScope is used to characterize the oblique helicoid when the cone angle $\theta$ is small. PolScope maps the optical retardance $\Gamma(x,y)$ of the sample, $\Gamma = \int \Delta n_{eff}\,dz$, where $\Delta n_{eff}$ is the effective birefringence of the heliconical state. For a small $\theta$, one can use an approximation $\Delta n_{eff} \approx \Delta n(1 - \frac{3}{2}\sin^2\theta)$ [7], where $\Delta n$ is the birefringence of the unwound $\hat{\mathbf{n}} = (1,0,0)$ state. As a measure of $\Delta n$, we use the experimentally determined birefringence of pure CB7CB, $\Delta n = 0.15$ at 106°C. This value yields $\Gamma = 1690$ nm for the unwound state in the cell of thickness $d = 11.2\,\mu m$, Fig. 1(e). When the field is reduced, the nematic-to-oblique helicoid transition is manifested by a cusp in the dependency $\Gamma(E)$ followed by a decrease of $\Gamma$. Such a behavior is expected because of the departure of $\theta$ from its zero value at $E \leq E_{NC}$, Eq.(3). The direct comparison of Eq.(3) and the experimental data in Fig.1(e) is difficult, as the director in the experiment is influenced by the boundary conditions; we resort to numerical simulations.

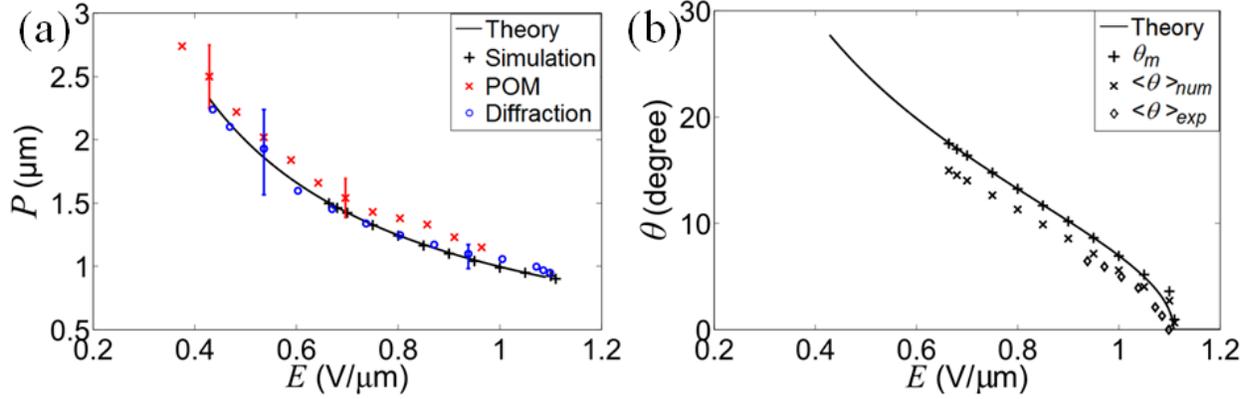

FIG. 2. Electric field dependence of (a) heliconical pitch; (b) cone angle, as deduced from the theory, numerical simulations and experiments.

**Director simulation.** Simulations of the director field are performed by minimizing the free energy functional based on bulk density (1) and surface anchoring, $F = \int f_b dV + \int f_s dS$, where surface anchoring for homeotropic alignment is represented by the Rapini-Papoular potential $f_s = \frac{1}{2}W\left[1-(\hat{\mathbf{z}}\cdot\hat{\mathbf{n}})^2\right]$; here $W$ is the anchoring coefficient and $\hat{\mathbf{z}}$ is a normal to the substrates. The simulations are performed for the experimentally relevant values $\kappa = K_3/K_2 = 0.12$, $P_0 = 7.5\,\mu m$, $E_{NC} = 1.1\,V/\mu m$, and assuming $W = 10^{-4}\,J/m^2$. The latter quantity does not influence the results much, as the typical contribution of the anchoring energy to the total energy is less than 0.1%. Minimization of $F$ reveals that the oblique helicoid is the main structural element of the system in the range $E_{N*C} \leq E \leq E_{NC}$, Fig.3(a). The effect of finite cell thickness and surface anchoring is in introducing the $z$-dependence of the polar $\alpha(z)$ and azimuthal $\psi(z)$ angles characterizing the orientation of the heliconical axis $\hat{\mathbf{h}} = (\sin\alpha\cos\psi, -\sin\alpha\sin\psi, \cos\alpha)$, and in $z$-dependence of the cone angle $\theta(z)$, Fig.3(b).

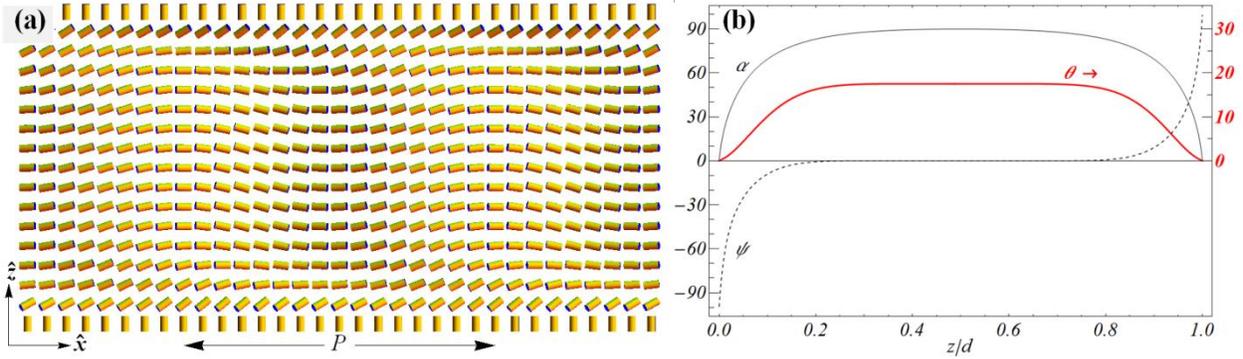

FIG.3. (a) The director field in the cell with the homeotropic anchoring under the in-plane electric field, simulated for parameters $\kappa = K_3/K_2 = 0.12$, $\kappa_1 = K_1/K_2 = 2.2$, $E/E_{NC} = 0.6$ and $d/P_0 = 1.49$, demonstrating a heliconical structure with $P = 0.20 P_0$. (b) In the heliconical state, the polar angle $\alpha$ (thin) and azimuthal angle $\psi$ (dashed) of the heliconical axis $\hat{\mathbf{h}}$ as well as the cone angle $\theta$ (thick) are functions of $z$ coordinate across the cell.

The field dependence of the pitch obtained in numerical simulations of the finite thickness cell reproduces the dependence $P \propto 1/E$ found experimentally and predicted by Eq.(2) very closely, Fig.2(a). The simulations also allow us to trace the field dependence of $\theta$. Since the surface anchoring makes the cone angle $z$-dependent, we distinguish two parameters: the value $\theta_m$ in the middle of the cell, and the average value $\langle\theta\rangle_{num} = \sin^{-1}\sqrt{\int \sin^2\theta(z) dz/d}$. As shown in Fig.3(b), $\theta_m$ follows closely the behavior expected analytically, Eq.(3). The average value $\langle\theta\rangle_{num}$ is somewhat smaller because of the surface anchoring effect, and matches well its experimental equivalent, determined as $\langle\theta\rangle_{exp} = \sin^{-1}\sqrt{\frac{2}{3}\left(1+\Gamma/(\Delta n \cdot d)\right)}$, Fig.2(b).

In conclusion, we demonstrate the electric field-controlled oblique helicoidal state of a cholesteric LC with positive dielectric anisotropy, predicted many decades ago, but so far not proven experimentally because of lack of materials with a sufficiently small bend-to-twist moduli ratio. In the explored cholesteric, based on molecular dimers, the bend elastic constant $K_3 = 0.3\,\text{pN}$ is determined to be about 10 times smaller than the twist constant, $K_2 = 2.6\,\text{pN}$, a very unusual result as compared to standard LCs formed by rod-like molecules. We establish the scenario of structural transitions in such a cholesteric, as the function of the applied electric field. At high fields, a uniaxial director parallel to the field is a stable state. At intermediate fields, the homochiral oblique helicoid forms with the axis parallel to the field. Both the pitch and the cone angle increase as the field is reduced. Finally, at low fields, the cholesteric right-angle helicoid emerges, with the axis perpendicular to the field.

Materials such as CB7CB have been connected to the so-called twist-bend nematic phase [11-15] that was recently shown by TEM observations to exhibit a nanoscale periodicity [7, 16] and the director field in the form of an oblique helicoid [7]. There are important differences between the oblique helicoidal structures induced by the electric field in the N* slab in this work and the ground state of the $N_{tb}$ phase. First, the period of the two structures is absolutely different, nanometers in $N_{tb}$ vs microns in N* (the latter can be controlled by the electric field and concentration of chiral additives). Second, because of the chiral dopant, the oblique helicoid in the N* case has the same chirality everywhere in the sample. In contrast, the $N_{tb}$ phase is formed by non-chiral molecules and thus should break into left-handed and right-handed domains [12]. As evidenced by TEM [7], these domains might be rather small, only 20-30 nm wide; generally, one expects the domain size to be dependent on the sample and its prehistory.

The observed homochiral oblique helicoidal state is close to the field-free chiral smectic C (SmC*) phase [1] and the phase observed in concentrated water suspensions of helical flagella isolated from Salmonella typhimurium [17]. In the latter case, the conical angle $\theta$ is fixed by the shape of relatively rigid flagellae. In the case of SmC*, $\theta$ is fixed by the molecular tilt within the smectic layers and the structure is modulated both in the sense of molecular orientation and materials density [1]. Absence of density modulation and the ability of the electric field to tune both the period and conical angle in the heliconical state of N* brings advantages in terms of proper alignment and electro-optical switching. Note also that each value of the pitch and cone angle is stable as long as the field is kept constant, which distinguishes the observed state from the transient structures that might occur in cholesteric cells when the field is abruptly removed [3]. One thus might expect that the electrically-controlled oblique helicoidal structure demonstrated in this work will find applications in devices such as tunable diffraction gratings, color filters, light deflectors and scatterers, wide-angle beam steerers, etc.

**Acknowledgements.** This work was supported by NSF DMR 1121288 (experiments, theory) and DOE Grant No. DE-FG02-06ER 46331 (numerical simulations). We thank Volodymyr Borshch, Young-Ki Kim, and Shuang Zhou for fruitful discussions.

* olavrent@kent.edu

**Supplemental material**

**Pitch measurement using Grandjean-Cano wedge**

A Grandjean-Cano wedge cell was produced by assembling two glass plates to form a wedge with an opening angle $\gamma = 0.6^o$. The uniform planar boundary conditions orient the N* helix perpendicularly to the cell walls. The number of half-pitches of N* increases as the wege gap increases. The domains with different number of half-pitches are separated by dislocation lines. The pitch is calculated as $P_0 = 2 \cdot S \cdot \tan\gamma$, where $S$ is the distance between neighboring dislocations. The temperature dependence of $P_0$ is shown in Fig.S1.

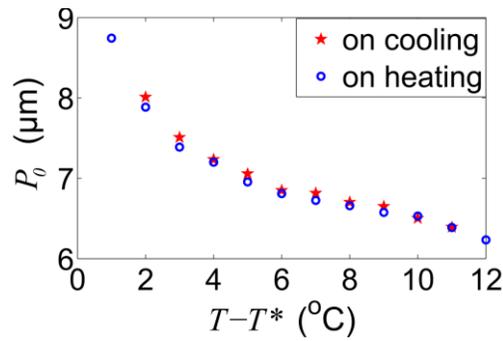

Fig.S1. Temperature dependent pitch $P_0$ of the chiral mixture CB7CB (99wt%)+S811 (1wt%)

**Electric field simulation**

Electric fields in the cells with an in-plane electric field applied with the help of either patterned ITO electrodes or by two aluminum foil electrodes are generally nonuniform. To calculate the field configuration, we used commercial finite-element modelling simulator, COMSOL Multiphysics. We used the experimentally known parameters of the cell, cell thickness 12μm, electrode gap 100-140μm, ITO thickness 0.15 μm, and glass plates thickness 1.1mm. The dielectric constant of glass is $\varepsilon_{glass} = 3.9$, and the dielectric constants of liquid crystal are $\varepsilon_\parallel = 7.3$ and $\varepsilon_\perp = 5.9$. As shown in Fig.S2, in the central part of the cell, the field is uniform and horizontal. For example, for the ITO case, the field is 0.7 V/μm with a 5% accuracy in the range $-20\,\mu m \leq x \leq 20\,\mu m$ and across the entire extension of the LC slab along the $z$-direction. For the aluminum electrodes, the field is 0.76V/μm with a 5% accuracy in the range $-20\,\mu m \leq x \leq 20\,\mu m$, Fig.S2(c,d).

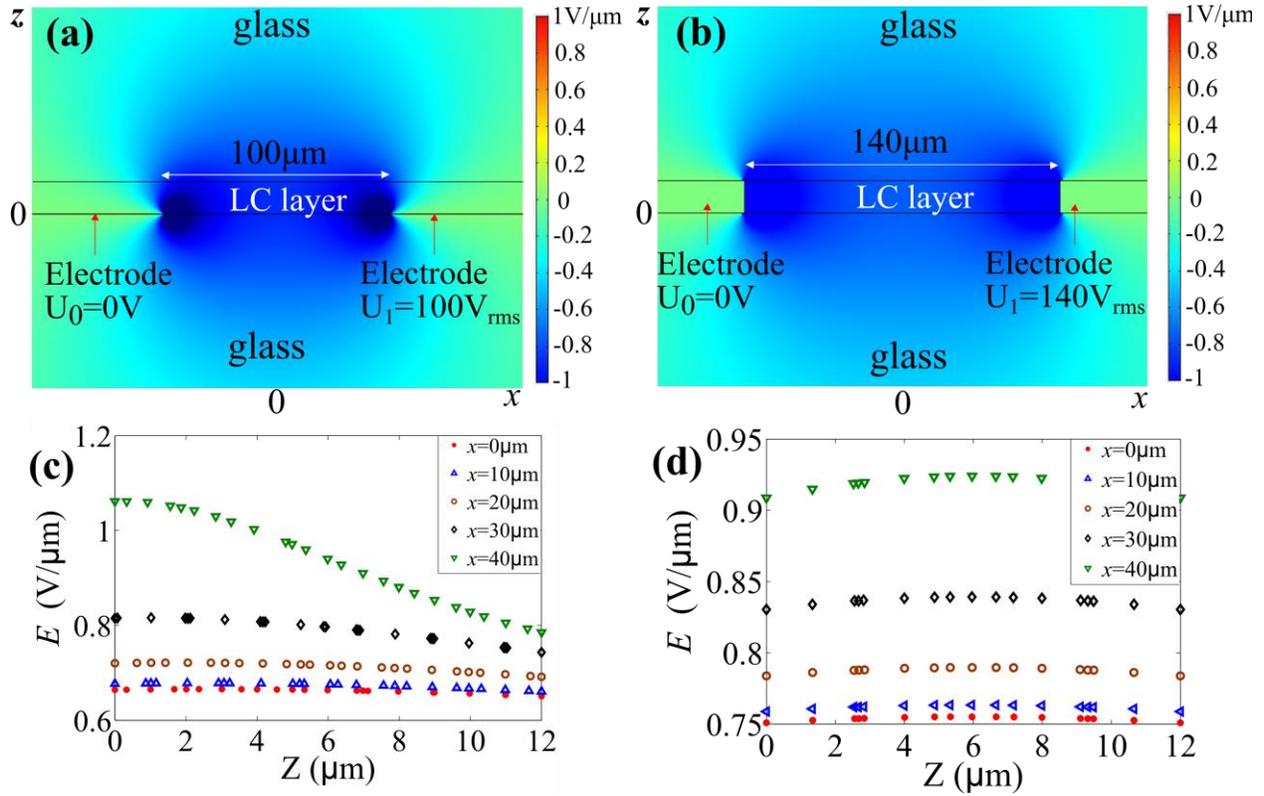

Fig.S2. Electric field in the in-plane switching cells: (a) patterned ITO electrodes at the bottom glass substrate with electrode gap 100μm, and applied voltage 100Vrms; (b) aluminum foil electrodes bewteen the glass substrates with electrode gap 140μm, and applied voltage 140Vrms. Electric fields strength in the patterned ITO cell (c) and aluminum electrodes cell (d), as a function of the $z$ coordinate, for different values of the $x$-coordinate.